\documentclass[aps,prd,twocolumn,showpacs]{revtex4}
\usepackage{graphicx}
\newcommand{\be}{\begin{equation}}
\newcommand{\ee}{\end{equation}}
\newcommand{\ben}{\begin{eqnarray}}
\newcommand{\een}{\end{eqnarray}}
\begin{document}

\title{Confining potential in a color dielectric medium
with parallel domain walls}
\author{D. Bazeia$^{a}$, F.A. Brito$^{a,b}$, W. Freire$^{a,c}$
and R.F. Ribeiro$^{a}$} \affiliation{$^a$Departamento de F\'\i
sica, Universidade Federal da Para\'\i ba,
\\
Caixa Postal 5008, 58051-970 Jo\~ao Pessoa, Para\'\i ba, Brazil
\\
$^b$Departamento de Ci\^encias F\'\i sicas e Biol\'ogicas and
$^c$Departamento de Matem\'atica
\\
Universidade Regional do Cariri, 63100-000 Crato, Cear\'a, Brazil}

\date{\today}

\begin{abstract}
We study quark confinement in a system of two parallel domain
walls interpolating different color dielectric media. We use the
phenomenological approach in which the confinement of quarks
appears considering the QCD vacuum as a color dielectric medium.
We explore this phenomenon in QCD$_2$, where the confinement of
the color flux between the domain walls manifests, in a scenario
where two 0-branes (representing external quark and antiquark) are
connected by a QCD string. We obtain solutions of the equations of
motion via first-order differential equations. We find a new color
confining potential that increases monotonically with the distance
between the domain walls.
\end{abstract}

\maketitle

%%%%%%%%%%%%%%%%%%%%%%%%%%%%%%%%%%%%%
\section{Introduction}

The color dielectric models have been used to describe
phenomenologically the confinement of quarks and gluons inside the
hadrons --- see Refs.~{\cite{fl,mit,slac,cornell}} for pioneering
papers on this subject. We shall investigate the quark confinement
in a system of two parallel domain walls separating different
color dielectric media. Such a system is regarded as a hadron and
our main goal in this paper is to search for a confining potential
of the color electric field. The color dielectric effect is
achieved via coupling between a color dielectric function $G$
chosen properly and the dynamical term of the gauge field. As one
knows the color vacuum in QCD has an analog in QED. In QED the
\textit{screening effect} creates an effective electric charge
that increases when the distance between a pair of
electron-antielectron decreases. On the other hand, in QCD there
exists an \textit{anti-screening effect} that creates an effective
color charge which decreases when the distance between a pair of
quark-antiquark decreases. This means that for small distance
(large momentum transfer) the quarks and gluons are considered
approximately free inside the hadrons (asymptotic freedom).

In the examples described above, the screening effect is
essentially due to the vacuum polarization in the relativistic
limit. However, it is believed to be possible to treat the physics
of quarks quite well in non-relativistic models of hadrons. This
is of special interest in physics of heavy quarks, where one
considers non-relativistic quarks and antiquarks connected by a
color flux tube known as a QCD string. The energy of such
configuration is described by confining potentials, i.e., by
potentials that increase linearly with the distance. In color
dielectric models one considers the QCD vacuum as a color
dielectric medium. This is somehow analog to what happens with two
electrical charges of opposite signs embedded in a polarizable
dielectric medium. This is a common phenomenon that occurs in
classic electrodynamics. The Lagrangian should contain a term like
$G\,F_{\mu\nu}F^{\mu\nu}$ in order to describe the dielectric
effects on the electric field, e.g., the displacement vector. The
dielectric function $G(r)$ usually assumes the behavior:
$G(r\!\gg\!d)\!\geq\!1$ and $G(r\!\ll\!d)\!<\!1$, where $d$ is the
diameter of the polarized molecules and $r$ represents an
arbitrary position on the dielectric medium. The QCD analog to
this electric effect is achieved using in the QCD Lagrangian the
term $G\,F^a_{\mu\nu}F_a^{\mu\nu}$. The color dielectric function
$G(r)$ in order to guarantee absolute color confinement must
assume the behavior: $G(r\!>\!R)\!=\!0$ in the color dielectric
medium (outside the hadron) and $G(r\!<\!R)\!=\!1$ inside the
hadron, where $R$ is the radius of the hadron.

For simplicity, in this paper we focus attention on QCD in
two-dimensions, QCD$_2$, for short. Two-dimensional QCD in the
large $N_c$ limit was introduced long ago \cite{thooft74}. It is a
truly relativistic field theory resembling the realistic four
dimensional QCD --- see, for instance, \cite{nefediev}. The theory
naturally exhibits confinement since the Coulomb force is
confining in two dimensions.

We choose a function $W(\phi,\chi)$ for the self-interacting part
of the real scalar fields $\phi$ and $\chi$. The function
$W(\phi,\chi)$ is chosen such that one of the scalar fields
provides kink (anti-kink) solutions and the other scalar field
contributes to the color dielectric function $G(\phi,\chi)$, with
the right behavior for confinement. We find a solution
representing two parallel domain walls that confine the color flux
in between the two walls. In $(3,1)$ dimensions, the kink
solutions are seen as thick domain walls which in the thin limit
are regarded as 2-branes that can be found in string/M-theory. In
the QCD realm, a 2-brane solution can be a place where the color
flux terminates. This possibility has been considered first in the
context of M-theory \cite{witten97} and also in field theory
\cite{trodden,campos,shifman,dvali,sakai}. In this scenario one
considers a QCD string (flux tube) ending on a 2-brane (quark or
antiquark). The QCD string itself is usually good to describe the
spectra of heavy mesons such as charmonium or bottomonium --- see
\cite{td,wil}. Since we shall restrict ourselves to a
two-dimensional theory, our kink (anti-kink) solutions will be
regarded as 0-branes (quark or antiquark), which are the ending
points for the QCD$_2$ string (\textit{flux line}). In our model
the two parallel domain walls solution (0-branes) that confines
the color field can be regarded as a QCD$_2$ string connecting a
quark to an antiquark.

In order to find classical solutions of the equations of motion we
take advantage of the first-order differential equations that
appear in a way similar to the Bogomol'nyi approach, although we
are not dealing with supersymmetry in this paper. The formalism
lead us to a suitable potential $V$ in order to make applications
feasible. The first-order equations for the scalar fields decouple
from the gauge field part. The dynamics of the fermionic sector,
used to describe quarks is not considered at this stage of the
calculations. Rather we regard the fermionic (quarks) effects via
an external color current source. We also assume that the set of
QCD$_2$ equations of motions can be replaced to another set, which
consider only Abelian fields. This is because through the color
dielectric function $G(\phi,\chi)$ the scalar field couples to an
average of the gauge field $A^a_\mu$ \cite{NPA448}. Thus, it
suffices to consider only the Abelian part of the non-Abelian
gauge dynamics.

We organize this work as follows. In Sec.~\ref{mod} we introduce
the Lagrangian of the QCD$_2$ in a color dielectric medium and
then we solve the equations of motion using first-order
differential equations. The parallel domain wall solution that we
find allows to have a confining color electric potential. In
Sec.~\ref{conc} we comment the results. Our notation is standard,
and we use dimensional units such that $\hbar\!=\!c\!=\!1$, and
metric tensor with signature $(+,-)$.

%%%%%%%%%%%%%%%%%%%%%%%%%%%%%%%%%%%%%%%%%%%%%%%%%%%%%
\section{QCD$_2$ with color dielectric function}
\label{mod}

A general theory of QCD$_2$ in a color dielectric medium can be
described by the Lagrangian \ben\label{QCD2} {\cal
L}^2&=&-\frac{1}{4}G(\phi,\chi)F^a_{\mu\nu}(x)
F_a^{\mu\nu}(x)\nonumber
\\
&&+\frac{1}{2}\partial_\mu\phi(x)\,\partial^\mu
\phi(x)+\frac{1}{2}\partial_\mu\chi(x)\,\partial^\mu
\chi(x)-V(\phi,\chi)\nonumber
\\
&&+{\bar\psi}(x)(\,i\,\gamma^\mu\partial_\mu-m)\psi(x)\nonumber
\\
&&+{\bar\psi}(x)( g\gamma^\mu A^a_\mu X_a-f(\phi,\chi)\,)\psi(x)
\een where $X_a$ are $SU(N_c)$ color matrices,
$F^a_{\mu\nu}\!=\!\partial_\mu A^a_\nu-\partial_\nu
A^a_\mu+g\,f^a_{bc}A^b_\mu A^c_\nu$, $x$ stands for $(x_0,x)$ and
we have suppressed the flavor indices. The quark spinor field
$\psi$ is in general coupled to the scalar fields via the Yukawa
coupling term ${\bar\psi}f(\phi,\chi){\psi}$. Here, however, we
first discard all the fermions, to focus on the bosonic background
fields. The presence of fermions will be further examined below,
to discuss how the walls can be charged.

In a medium which accounts mainly for one-gluon exchange, the
gluon field equations linearize and are formally identical to the
Maxwell equations \cite{wil}. In this sense, it suffices to
consider only the Abelian part of the non-Abelian strength field
\cite{NPA448}, i.e., $F^a_{\mu\nu}\!=\!\partial_\mu
A^a_\nu-\partial_\nu A^a_\mu$. Further, without loss of generality
we can suppress the color index $``a"$ if we take an Abelian
external color source \cite{slusa} due to quarks or antiquarks
represented by an external color current density $j^\mu_a$. This
is because the results both for Abelian and non-Abelian cases are
very similar \cite{slusa,dick1,dick2,cha1,cha2}. We account for
all these facts, thus the model we start with to investigate
confinement is the effectively Abelian Lagrangian \ben\label{m}
{\cal
L}^2_{eff}\!&=&\!-\frac{1}{4}G(\phi,\chi)F_{\mu\nu}F^{\mu\nu}+j_\mu
A^\mu\nonumber\\
&+&\frac{1}{2}\partial_\mu\phi\,\partial^\mu\phi+
\frac{1}{2}\partial_\mu\chi\,\partial^\mu\chi-V \een where $j_\mu$
is an Abelian external current density.

%%%%%%%%%%%%%%%%%%%%%%%%%%%%%%%%%%%%%%%%%%%%%%%%%
\subsection{Equations of motion}

We search for static solutions of the equations of motion. Thus,
we consider the fields depending only on the spatial coordinate,
$x$. The equations of motion that follow from (\ref{m}) are
written as \ben \frac{d^2\phi}{dx^2}&=&V_{\phi} -
\frac{1}{2}\,G_{\phi}\left(\frac{dU}{dx}\right)^2
\\
\frac{d^2\chi}{dx^2}&=&V_{\chi} -
\frac{1}{2}\,G_{\chi}\left(\frac{dU}{dx}\right)^2 \een and \be
\frac{d}{dx}\left[G(\phi,\chi)\frac{dU}{dx}\right]=\rho(x) \ee
where we have set $U\!=\!A^0$ and $\rho\!=\!j^0$. Also, $V_\phi$
and $G_\phi$ stand for derivatives with respect to $\phi$, and so
forth. Now, we choose the potential in the form \ben\label{p}
V(\phi,\chi,x)=\frac{1}{2}W^2_{\phi}+\frac{1}{2}W^2_{\chi}-\frac{1}{2G}
\left[\int^x dx'\,\rho(x')\right]^2 \een where the function
$W(\phi,\chi)$ is smooth everywhere, in general. The form of
$V(\phi,\chi,x)$ in (\ref{p}) appears because we want to use
first-order differential equations, instead of the second-order
equations of motion. To see this, we notice that the above
equations of motion are solved by $\phi(x),\,\chi(x)$, and $U(x)$
that solve the new set of first-order equations \ben\label{edo1}
\frac{d\phi}{dx}&=&W_{\phi}
\\
\label{edo2} \frac{d\chi}{dx}&=&W_{\chi}
\\
\label{edo3} G(\phi,\chi)\frac{dU}{dx}&=&\int^x dx'\,\rho(x') \een
This result is inspired in the procedure introduced in
Ref.~{\cite{bog}}, although here the situation is quite different
because we are not minimizing the total energy of the system --
see, for instance, Ref.~{\cite{bb2000}} for further details.
However, we can show that the above first-order equations solve
the equations of motion.

To solve the first-order equations we first notice that the
equations (\ref{edo1}) and (\ref{edo2}) do not couple with the
other equation (\ref{edo3}). We take advantage of this and we
define the function $W(\phi,\chi)$ to look for solutions of
(\ref{edo1}) and (\ref{edo2}). We consider
 \be\label{func01}
W(\phi,\chi)=\lambda\,a^2\phi-\frac{\lambda}{3}\,\phi^3-\mu\,\phi\chi^2
 \ee
where $\lambda,\mu$ and $a\!>\!0$ are real parameters. For this
particular $W$, the dynamical system (\ref{edo1}) and (\ref{edo2})
has four singular points, given by $(\phi\!=\!\pm a,\chi\!=\!0)$
and $(\phi\!=\!0,\chi\!=\!\pm\sqrt{\lambda/\mu}\,a)$, for
$\lambda/\mu>0$, which are interpolated by domain wall solutions
\cite{bb2000,bsr,brs,bb,shvol,bbb}. Before going into this, let us
now specify the functions $\rho(x)$ and $G(\phi,\chi)$. Since we
are representing here quarks and antiquarks as domain walls
(0-branes) with non-zero width, the color charge density is not
necessarily point-like. In this way we introduce the color charge
distribution on the walls as
 \be\label{func1}
\rho(x\pm s)=q\frac{\eta}{2}\,{\rm sech}^2\eta(x\pm s)
 \ee
where $s$ is the {\it center} of the distribution, which we choose
to be real and positive.  This charge density is interesting since
it tends to be point-like $[q\delta(x\pm s)]$ in the limit as
$\eta\to\infty$. The point-like limit is realistic as long as
$x\!\gg\!\Delta$, where the {\it width} $\Delta\!\sim\!1/\eta$ is
of the same order as the quark (or antiquark) radius. As we show
below, $\eta$ is defined according to the width of the domain wall
solutions that solve (\ref{edo1}) and (\ref{edo2}).

Let us now discuss the choice (\ref{func1}) that we have just done
for the color charge density on the walls. It can be justified due
to the presence of fermion zero modes on the walls. To show this
explicitly, we consider the Lagrangian density ${\cal
L}^{\prime}\!=\!{\cal L}^{\prime}_{eff}+{\cal L}_{D}+{\cal
L}_{Y}$, where ${\cal L}^{\prime}_{eff}$ is given in (\ref{m}),
with $j_\mu\!=\!q\bar{\psi}\gamma_\mu\psi$. The other fermion
contributions are given by the Dirac Lagrangian density ${\cal
L}_{D}$ and the Yukawa coupling term ${\cal L}_{Y}$. The variation
$\delta{\cal L}^{\prime}/\delta\bar{\psi}=0$ gives the fermion
equation of motion
 \ben
\label{ferm} i\gamma^\mu\partial_\mu\psi+q\gamma^\mu A_\mu \psi
-[m+f(\phi,\chi)]\psi=0.
 \een
We choose $f(\phi,\chi)\!=-m+k\Phi(x)\!$, with
$\Phi(x)\!=\!\phi(x)-\phi_0$ where $\phi(x)$ is the bosonic
background solution for the scalar field $\phi$, which is given
below, in Eq.~(\ref{sol11}). The meaning of $\phi_0$ is going to
be clear later. Looking for two-dimensional solutions
($\mu\!=\!t,x$) we start with the Ansatz
 \ben \label{atz}
\psi=e^{iqA_0(x)t}h(x)\epsilon_\pm
 \een
where $\epsilon_\pm$ is a constant spinor and $A_0(x)$ is the
color electric potential. Substituting (\ref{atz}) in (\ref{ferm})
and using the fact that $\gamma^x\epsilon_\pm=\pm i\epsilon_\pm$,
we find the solution
$h(x)\!=\!C\,\exp{[-iqA_0(x)t]}\times\exp{[\mp
k\int^x\Phi(x')dx']}$, where $C$ is an arbitrary constant. Now we
use Eq.~({\ref{atz}}) to get to the spinor solution
 \ben\label{spsol}
 \psi(x)=\Big[C\,e^{\mp k\int^x\Phi(x')dx'}
\Big]\epsilon_\pm
 \een
The solution (\ref{sol11}) [together with (\ref{sol12})] represent
two parallel domain walls. Each domain wall can be treated
separately. Explicitly, this can be done by shifting $x$ in
(\ref{sol11}) as $x\to\pm s+2z$ such that one has
\ben\label{phip0}\Phi(z)=\frac{a}{2}\Big[\tanh{\frac{2(z\pm s)}
{\Delta}}+\tanh{\frac{2z}{\Delta}}\Big]-\phi_0. \een Now it is
clear that $\phi_0$ must be $(a/2)\tanh{({2z}/\Delta)}$ such that
we are able to study each domain wall isolated which are described
as
 \ben \label{phip}
\Phi(z)=\frac{a}{2}\tanh{\frac{2(z\pm s)}{\Delta}}.
 \een
Note that, locally, each domain wall solution (\ref{phip}) changes
its sign in $z\!=\!-s$ and $z\!=\!s$. Thus, there are fermion zero
modes \cite{jack} localized on the domain walls at the points
$x\!=\!\pm s$. In fact, we can find the zero mode solutions
explicitly: substituting (\ref{phip}) in (\ref{spsol}) and
integrating in $x'\,\!(=\!z)$ we find the normalizable chiral zero
modes
 \ben
\label{zmode} \psi=C\,{\rm sech}{\frac{2(x\pm
s)}{\Delta}}\epsilon_\pm
 \een
where we have set $k=4/a\Delta$. With these solutions, the
localized charge on the domain walls due to fermionic charge
carriers is given by using the current density
$j_\mu\!=\!q\bar{\psi}\gamma_\mu\psi$. The charge density is
$\rho\!=\!j_0\!=\!q\psi^\dagger\psi$ and so, we use the fermion
solutions (\ref{zmode}) to find $\rho(x\pm s )=q(\eta/2)\,{\rm
sech}^2{\eta(x\pm s)}$, where we have identified
$\eta\!=\!2/\Delta$ and $\eta/2=C^2R$, with
$R\!=\!\epsilon^\dagger_\pm\epsilon_\pm$ being a constant number.
This charge distribution is the charge distribution that we have
chosen in Eq.~(\ref{func1}).

On the other hand, we define the color dielectric function as
 \be\label{func3}
G(\phi,\chi)=A\,\chi^2 \ee where $A$ is a normalization constant,
dimensionless. As we show below, this definition is sufficient to
produce an absolute color confining effect since the $\chi^2$
solution has the appropriate behavior, to make the above $G$ to
have the required asymptotic profile, as we have mentioned in the
introduction.

We notice that the limit $\mu\to0$ decouples the scalar fields,
making the model meaningless. Also, the choice (\ref{func01})
appears as a simple extension of the $\phi^4$ model (with
spontaneous symmetry breaking) to include another field, the
$\chi$ field, which interacts with the gauge field via the color
dielectric model that we propose, inspired in the former
Refs.~{\cite{wil,slusa,dick1,dick2}}.

%%%%%%%%%%%%%%%%%%%%%%%%%%%%%%%%%%%%%%%%%%%%%%%%
\subsection{Confinement with two parallel domain walls}

Among several domain wall solutions connecting the four singular
points of the system (\ref{edo1}) and (\ref{edo2}), when
$W(\phi,\chi)$ is given as in Eq.~(\ref{func01}), there is a
particular solution of interest here: the two parallel domain wall
solution reported in Refs.~{\cite{shvol,volosh,gani}}. To see this
explicitly, we substitute (\ref{func01}) into the first-order
equations (\ref{edo1}) and (\ref{edo2}) to obtain
 \ben\label{m1}
\frac{d\phi}{dx}&=&\lambda(a^2-\phi^2)-\mu\chi^2
\\
\frac{d\chi}{dx}&=&-2\mu\phi\chi\label{m2}
 \een
This system presents analytical solutions such as
\ben\label{sol11} \phi&=&\frac{a}{2}\left[\tanh{\frac{\lambda
a(x+s)}{2}}+\tanh{\frac{\lambda a(x-s)}{2}}\right]
\\\label{sol12}
\chi&=&\pm\sqrt{2}\,a\left[1-\tanh{\frac{\lambda
a(x+s)}{2}}\tanh{\frac{\lambda a(x-s)}{2}}\right]^{1/2} \een if
one sets $\mu\!=\!\lambda/4$. This solution clearly represents two
parallel domain walls at the positions $x\!=\!-s$ and $x\!=\!s$.
It provides a color dielectric function $G=A\,\chi^2$ (with
$A\!=\!1/2a^2$) with the proper behavior inside and outside the
region in between the domain walls: $G\!\to\!1$ for $-s\!\leq
x\!\leq\!s$, and $G\!\to\!0$ for $x\!>\!s,\,x\!<\!-s$, as we
illustrate in Fig.~1. This characterizes an absolute color
confinement, as desired. The interpretation of this is that the
color flux inside such a region connects charged domain walls with
opposite color charges.

%%%%%%%%%%%%%%%%%%%%%%%%%%%%% FIGURE %%%%%%%%%%%%%%%%%%%%%%%%%%%%%%%%
\begin{figure}[h]
\includegraphics[{height=5.0cm,width=7.0cm}]{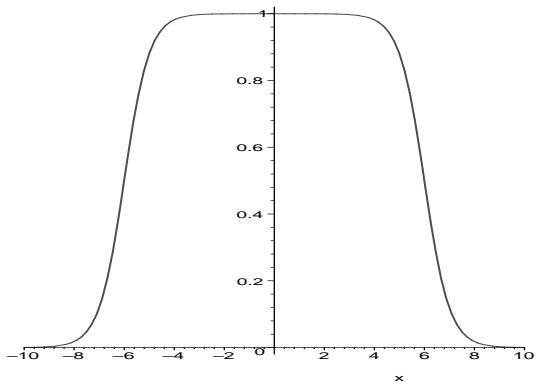} \caption{The color
dielectric function behavior is plotted in arbitrary units. It
shows that the color field is confined in between the two domain
walls, which are centered at $x=\pm s=\pm6$.}\label{fig1}
\end{figure}
%%%%%%%%%%%%%%%%%%%%%%%%%%%%%%%%%%%%%%%%%%%%%%%%%%%%%%%%%%%%%%%%%%%%%%%

It is now clear that $\eta$ in the color charge density function
(\ref{func1}) should agree with the domain walls width that
appears in the solutions (\ref{sol11}) and (\ref{sol12}), i.e.,
$\eta\!=\!\lambda a/2$. Let us now write the equation
(\ref{edo3}), with $\rho(x)$ and $G(\phi,\chi)$ defined as in
(\ref{func1}) and (\ref{func3}) respectively, in the form
 \ben\label{de}
\label{func02}\chi^2\,\frac{dU}{dx}=q
a^2\left(\tanh{\frac{x+s}{\Delta}}-
\tanh{\frac{x-s}{\Delta}}\right)
 \een
where we are using $\Delta=2/\lambda a$ to represent the width of
each domain wall. This expression has to be integrated over the
external color charge density of the quark-antiquark pair
represented by the solution obtained in Eqs.~(\ref{sol11}) and
(\ref{sol12}). The convention is such that the lines of the color
electric flux come from the domain wall on $x\!=\!-s$ (a quark)
and end on the domain wall on $x\!=\!s$ (an antiquark).

We substitute the solution (\ref{sol12}) into the equation
(\ref{func02}) to perform the integration in the interval that
goes from $-s-\Delta/2$ to $s+\Delta/2$, which is the region where
the solutions (\ref{sol11}) and (\ref{sol12}) appreciably deviate
from trivial vacuum states. The result is
 \be\label{sol21}
U(s)=2\sigma\left(s+\frac{\Delta}{2}\right)\,\tanh(2s/\Delta)
 \ee
where we are using $\sigma=q/2$, as the string tension of the
string that connects the quark-antiquark pair. At large distance,
for  $s\!\gg\!\Delta$ we obtain
 \be
U(s)\approx\sigma\,\Delta+2\,\sigma s,
 \ee
which nicely reproduces the well-known linear behavior in quark
confinement This is the scenario we have in the thin-wall limit.

The above scenario is introduced to represent a meson. In the
region outside the parallel domain walls (the vacuum) the field
$\chi\!\to\!0$ and so does the color dielectric function $G$,
whereas the field $\phi$ is non zero. This scalar field can be
excited around its vacuum $\phi_0$. Such excitation can be
identified with $glueballs$, i.e., pure gluonic hadrons
surrounding the meson. In our model the glueball mass is given by
$\partial^2 V(\phi)/\partial\phi^2|_{\phi=\phi_0}$, where
$V(\phi)\!=\!(1/2)[W_{\phi}^2+W_{\chi}^2]|_{\chi=0}$ with the
$W(\phi,\chi)$ chosen as in (\ref{func01}), with $\mu=\lambda/4$.
The glueball mass is then given by $m_{GB}^2\!=\!4\lambda^2 a^2$.
One can redefine some quantities in terms of the glueball mass;
for instance, the scale of symmetry breaking can be given as
$a^2\!=\!3m_q/m_{GB}$, once we are identifying $m_q\!=\!2\lambda
a^3/3$ (the mass $m_q$ of each quark is identified with the
tension of each isolated wall). For heavy quarks we have
$m_q\!\gg\!m_{GB}$ and then $a^2\gg1$. Since the wall width
$\Delta$ is $(2/\lambda a)$ and so, for finite $\lambda$ the
thin-wall limit is then ensured, in accordance with the previous
results.

Let us now turn attention to the small distance ($s\!\ll\!\Delta$)
behavior of our model. We cannot use the above result
(\ref{sol21}) to examine the small distance behavior of the
potential. This limitation is to be expected, since at small
distance the two walls overlap and change the scenario. To
circumvent this issue we turn back to solutions (\ref{sol11}) and
(\ref{sol12}), to see that the limit $s\ll\Delta$ leads to other
solutions, as follows \ben \label{nsol1}
\phi(x)&\to&\phi^{(0)}(x)=a\, \tanh{\left(\frac{x}{\Delta}\right)}
\\
\label{nsol2} \chi(x)&\to&\chi^{(0)}(x)=\pm\sqrt{2}\,a\,{\rm
sech}{\left(\frac{x}{\Delta} \right)} \een The solutions
$\phi^{(0)}$ and $\chi^{(0)}$ were first obtained in
Ref.~{\cite{bsr}}; they represent genuine two-field solutions, not
two-wall solutions anymore. This represents the thick wall
solution with internal structure \cite{brs,bb}. The fields $\phi$
and $\chi$ play another game now. The field $\phi$ is regarded to
describe a ``bag'' \cite{mit,slac} filled with gluons which is
represented by $\chi$. It is not difficult to see that in doing
all the above calculations in order to obtain the color potential
using the solutions (\ref{nsol1}) and (\ref{nsol2}) we end up with
a new confining potential.  The leading contribution that appear
from the right hand side of Eq.~(\ref{de}) in the limit
$s\!\ll\!\Delta$ vanishes, making the potential constant,
describing absence of force in between the gluons inside the bag,
in agreement with the asymptotic freedom behavior of QCD. The bag
represents a purely gluonic hadron, i.e., a {\it gluewall}, in
analogy to the three-dimensional glueball. Our understanding is
that at very small distance the pair quark-antiquark annihilates,
decaying into a colorless gluewall.

%%%%%%%%%%%%%%%%%%%%%%%%%%%%%%%%%%%%%%%%%%%%%%%%%%
\section{Comments and conclusions}
\label{conc}

In this paper we have presented a scenario which can describe
quark confinement via a monotonically increasing color electric
potential. This is achieved by considering a color flux confined
to a region in between two parallel domain walls. In
two-dimensional QCD, the color flux is confined to a {\it flux
line} ending on 0-branes. This is similar to QCD strings
connecting 2-branes in string/M-theory \cite{witten97}. The color
electric potential varies linearly with distance, in the large
distance limit between the domain walls, in the thin-wall limit.
On the other hand, at very small distance, our results predict
that the pair quark-antiquark annihilates to form a colorless
gluewall. This is the (analog of a) purely gluonic hadron
described by a bag represented by a domain wall with internal
structure.

Despite its simplicity, we see that our model presents a very
reasonable description of quark confinement at relatively large
distance. This result encourages us to investigate more realistic
models, taking into account mainly the non-Abelian character of
the color dielectric medium. Another line of investigation is
directly related to the way we couple the gauge field to the
scalar fields via the dielectric function $G(\phi,\chi)$. This
coupling is somehow similar to the coupling that appears in
Kaluza-Klein compactifications involving non-Abelian
\cite{dick1,dick2} and Abelian \cite{mts} gauge fields. In this
sense, an issue to be pursued refers to the relation between the
two scalar fields that appear in the present work, and the dilaton
and moduli fields used in Ref.~{\cite{mts}}.

%%%%%%%%%%%%%%%%%%%%%%%%%%%
\acknowledgments

We would like to thank C. Pires for useful discussions, and CAPES,
CNPq, PROCAD and PRONEX for partial support. FB and WF would like
to thank Departamento de F\'\i sica, UFPB, for hospitality. WF
also thanks FUNCAP for a fellowship.

%%%%%%%%%%%%%%%%%%%%%%%%%%%%%%%%%%%%%%%%%%%%%%%%%%%%%%%%%%%%%%%%%%%%

\end{document}